\documentclass[twocolumn,APS]{revtex4}

\usepackage{graphicx}
\usepackage{amsmath}
\usepackage{epsf}

\begin{document}

\preprint{version LDC \today}

\title{Inversion of two-band superconductivity at the critical electron doping of (Mg,Al)B$_2$}

\author{L. D. Cooley, A. J. Zambano, A. R. Moodenbaugh, R. F. Klie, Jin-Cheng Zheng, and Yimei Zhu}
  \affiliation{Brookhaven National Laboratory, Upton, NY 11973 USA}

\date{\today}

\begin{abstract}
Electron energy-loss spectroscopy (EELS) was combined with heat capacity measurements to follow the change of superconductivity with systematic Al doping of MgB$_2$.  By using x-ray diffraction and Vegard's law to assess the actual Al content in the samples, changes in behavior were found to be much more in agreement with theoretical predictions than in earlier studies.  EELS data show that $\sigma$-band hole states disappear above 33\% Al, approximately the composition at which the $\sigma$ band Fermi surface is predicted to lose its cylindrical shape in reciprocal space and break apart into ellipsoidal pockets.  At this composition, the $\sigma$ gap obtained from the heat capacity data falls to the level of the $\pi$ gap, implying that band filling results in the loss of strong superconductivity on the $\sigma$ band.  However, superconductivity is not quenched completely, but persists with $T_c < 7$~K up to about 55\% Al, the Al concentration at which the entire $\sigma$ band is predicted to fall below the Fermi surface.  Since, in the region $0.33 \alt x \alt 0.55$, only the $\pi$ band has appreciable density of states, it becomes the stronger of the 2 bands, thus inverting the 2-band hierarchy of MgB$_2$.  
\end{abstract}


\maketitle

%
%
%
%

Although the two-band superconductivity was first studied theoretically in 1959 \cite{Suhl},  its detailed experimental study has not been possible until the discovery of superconductivity near 40 K in MgB$_2$ in 2001 \cite{Akimitsu}.   The high critical temperature $T_c$ results from two key ingredients: First, the boron $\sigma$ band, which is deep beneath the Fermi energy $E_F$ in graphite, sits just above the Fermi energy and provides a high density of $p_{xy}$ hole states \cite{Mazin, Choi}.  Second, the boron-boron $E_{2g}$ bond oscillation is renormalized downward in energy (softened), into the range where the phonon coupling to the hole states is nearly perfect \cite{Ahn, Yildirim}.  The 2-dimensional $\sigma$ band carries along with it superconductivity in a 3-dimensional $\pi$ band with $p_z$ electron states, which has a smaller superconducting gap energy and weaker electron-phonon coupling \cite{Choi}.  

Since the $\sigma$ band appears to be so crucial to superconductivity, many experiments have attacked the central question of how superconductivity changes upon substituting Al or C for Mg or B, respectively, thereby providing extra electrons that fill its hole states.  A common observation has been that such dopants destroy superconductivity \cite{Slusky, Postarino, Kazakov, Ribeiro, Wilke}.  For Al doping, $T_c(x)$ falls with increasing Al concentration $x$, sometimes increasing slope at $x \approx 0.3$ and reaching zero for $x = 0.5$ to 0.6 \cite{Postarino, Luo, Li-JQ}.  The loss of superconductivity at $x \approx 0.5$ correlates with changes seen in NMR data \cite{Serventi, Papavassiliou}, optical reflectance \cite{Yang}, and infrared absorption \cite{Postarino}.  Heat capacity measurements \cite{Putti} show an approximately linear decrease of both gaps out to $x = 0.4$, suggesting that interband scattering does not affect these trends strongly.  A rise of the Fermi level consistent with band filling has been inferred from x-ray absorption spectroscopy (XAS) \cite{Yang}, but quantitative data is not available.

Theoretical predictions are generally consistent with these observations, but also add unusual details.  Electronic structure calculations within the virtual crystal approximation predict that the cylindrical Fermi surfaces attributed to the boron $\sigma$ band collapse at the $\Gamma$ point in reciprocal space when $x \approx 0.3$, separating the cylinders into strings of ellipsoidal pockets \cite{delaPena}.  Complete filling of the $\sigma$ band holes was predicted to occur at $x = 0.56$.  The topological change at $x = 0.3$ suggests abrupt changes in the properties occur for somewhat lower Al content than observed in experiments.  One such change predicted to occur within the 2-band Eliashberg formalism \cite{Bussman-Holder,Ummarino} is a crossover in the hierarchy of superconducting gap energies for the two bands, with the $\pi$ band becoming stronger than the $\sigma$ band at high doping, opposite to the order for pure MgB$_2$.  In this region, the predicted $T_c$ is also much lower than observed experimentally, perhaps due to uncertainty of the true amount of Al actually incorporated in the superconductor or the inhomogeneous distribution of Al.

In this Letter we present experimental data that validates these theoretical predictions better than in earlier work, including the inversion of the relative strength of superconductivity on the 2 bands when the Al content is high.  We combine spectroscopy, which tracks the loss of hole states on the $\sigma$ band, with heat capacity measurements, which provides information about the behavior of both gaps.  We also analyze data in terms of the relative change of the unit cell volume $v(x)=V(x)/V_0$, where $V_0 = V(0)$ is the unit-cell volume for pure MgB$_2$, and $V(x)$ is the unit-cell volume obtained from x-ray diffraction data after reaction.  Since the crystal lattice changes systematically with added Al, this procedure gives a better volumetric representation of the Al content actually present, on average, in the superconducting phase.  Thus, we can assert that the behavior that is observed is intrinsic to Al-doped MgB$_2$.  

\begin{figure}
\includegraphics{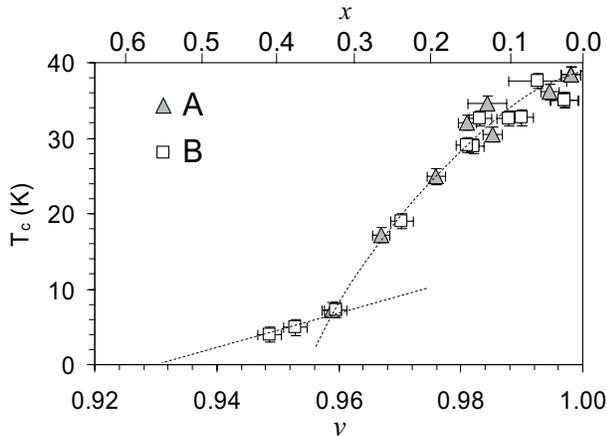}
\caption{Critical temperature as a function of the normalized volume of the unit cell taken from x-ray diffraction data for samples made by a short reaction (A) and a long, hot reaction (B).  The curves are guides to the eye.  The axis along the top edge indicates Al concentration calculated for a simple rule of mixtures between MgB$_2$ and AlB$_2$ (Vegard's law).}
\label{f1}
\end{figure}

Two sets of samples were prepared using reactions at opposite limits of what is routinely used in the literature in terms of time and temperature: a short reaction ``A'', 850 $^\circ$C for 1 hr, which was stopped just short of full consumption of reactants (thus the superconducting phase is ``as formed''); and a very hot, long reaction ``B'', 1200 $^\circ$C ramped down to 700 $^\circ$C over $>80$ hours like that in \cite{Badr}.  Full details of the structural characterization will be described in more detail elsewhere.  It is important to note here that the crystal lattice parameters for sample set B generally obeyed a simple mixing rule between MgB$_2$ and AlB$_2$ endpoints (Vegard's law) and the composition measured after reaction by energy-dispersive x-ray spectroscopy in a scanning electron microscope.  These samples thus appeared to be homogeneous in composition.  Samples from set A, on the other hand, generally incorporated Al more slowly than Mg, and also displayed other signs of inhomogeneity.  The x-ray diffraction characterization did not indicate the presence of any Mg-Al ordering \cite{Zandbergen} for sample set B, while broadening of the (002) peaks was evident for set A for $0.15 \le x \le 0.25$.  

Despite the evidence for inhomogeneity in one sample set, superconducting properties analyzed in terms of $v$ for both sets of samples fell onto a single curve, as for $T_c(v)$ shown in fig.~\ref{f1}.  Uncertainty bars on this plot represent the 10 to 90\% width of the $T_c$ transition in a 1 mT background field, measured after cooling in zero field.  The $T_c(v)$ curve exhibits a steep fall for $v > 0.96$, down to a critical temperature of about 7 K.  For $v < 0.96$ the inductive transitions clearly showed bulk supercondcutivity, where in this region $T_c$ extrapolates to 0 at $v \approx 0.93$.

The large grain size produced by reaction B provided excellent samples to probe the electronic structure with electron energy loss spectroscopy (EELS) \cite{Egerton, Browning}.  We used a field-emission scanning transmission electron microscope (STEM) system used previously to probe the electronic structure of individual grains of pure MgB$_2$ \cite{Klie}.  EELS reduces the effects of surface contamination because the probe beam passes through the crystal volume.  Furthermore, special crystal orientations can be selected easily in the large thin area of a polycrystalline sample.  Good correspondence between EELS and XAS was obtained in a study of the hole states in pure MgB$_2$ \cite{Zhu}.  In the present experiment, we extend those analyses to Al doped samples from set B.  

The different natures of the electronic $p_{xy}$ states and $p_z$ states, which are associated with the $\sigma$ band and the $\pi$ band respectively, lends a key advantage to the spectroscopy experiment.   In {\em pure} MgB$_2$, the $p_{xy}$ density of states (DOS) has a steep edge as a function of energy, extending from several eV below $E_F$ to about 0.9 eV above $E_F$ \cite{Zhu}.   The unoccupied $p_{xy}$ states farther above $E_F$ are few in number for about 5 eV, where a sharp rise in the $p_{xy}$ DOS marks the main boron $K$ edge.  In contrast to these sharp features of the $\sigma$ band, the $p_z$ DOS is essentially constant with energy over that same range, providing a relatively constant background to the spectroscopy measurements.  These differences are amplified further by using a large collection aperture and aligning the electron beam normal to the boron planes, which produces measured intensity that is dominated by the momentum transfer into $p_{xy}$ states \cite{Klie}.  

\begin{figure}[!tb]
\includegraphics{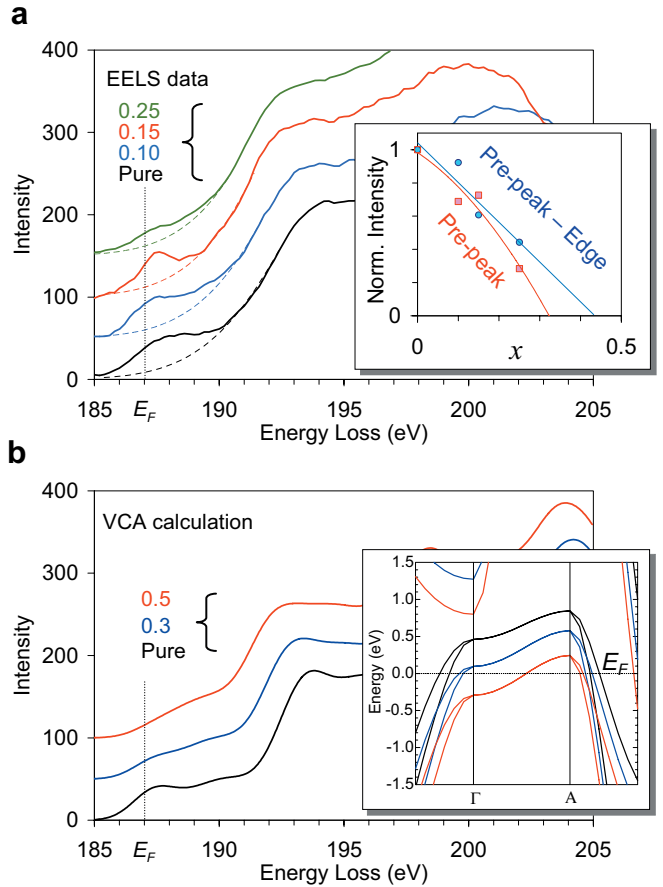}
\caption{STEM-EELS data are shown in plot (a).  The dashed line indicates a Gaussian fit to peaks at higher energy, which was subtracted to estimate the pre-peak intensity.  The inset shows the normalized integrated intensity of the pre-peak at 186--189 eV.  Plot (b) shows simulated EELS spectra obtained by density-functional theory using the virtual crystal approximation.  The inset shows the calculated band structure for the important $\sigma$ band.  In both plots (a) and (b), the Fermi energy is fixed and the curves are offset in intensity for clarity.  }
\label{f2}
\end{figure}

The essential features of the EELS data, shown in fig.~\ref{f2}a, are a pre-peak at 186--189~eV energy loss followed by a strong series of peaks at 190 eV and above.  Here, the spectra are aligned at 187 eV, which represents $E_F$.  The pre-peak is derived from core electrons being excited into the unoccupied $p_{xy}$ states at the top of the $\sigma$ band.  As Al is added, a systematic loss of pre-peak intensity occurs, indicating the rise of $E_F$ into the flat portion of the $\sigma$ band.  This behavior is very similar to that measured by XAS \cite{Yang}, which included the intensity due to the $p_z$ states.  The main $K$ edge also shifts downward, which will be described in a separate paper.  A quantitative assessment of the $p_{xy}$ DOS in this region as a function of doping was made by integrating the pre-peak intensity and normalizing to the integrated intensity for $x = 0$.  The inset of fig.~\ref{f2}a shows the 2 different approaches used: First, a 2 eV wide Gaussian curve was fit to the peak intensity at approximately 187.5 eV, the width being due to the broadening of the STEM.  Second, the tails of main edge peaks at $\sim 192$ and $\sim 195$~eV were fit by a 6 eV wide Gaussian (dashed curves in fig.~\ref{f2}a) and subtracted, and the remaining intensity was integrated from 185 to 191 eV.  As shown in the inset to fig.~\ref{f2}a, extrapolating either integrated intensity gives the Al content range where the $\sigma$ band DOS becomes too small to contribute appreciable intensity, which evidently occurs for $x \ge 0.32$.  

The STEM-EELS data are in good agreement with density-functional calculations based on the virtual crystal approximation, shown in fig.~\ref{f2}b.  Here, the DOS calculations have been smoothed to simulate the broadening in the experiment.  Fig.~\ref{f2}b shows that there is no pre-peak intensity for $x = 0.30$, at which point $E_F$ intersects the $\sigma$ band at the $\Gamma$ point (fig.~\ref{f2}b inset).  The remaining flat portion below 191 eV is due to the small amount of $p_z$ states that are sampled by the simulated collection conditions, as indicated by the lack of any change in this region between $x = 0.30$ and $x = 0.50$.  The overall calculations gave results very similar to those reported in \cite{delaPena}, including the position of $E_F$ relative to the bands as shown in the inset of fig.~\ref{f2}b.

\begin{figure}[!tb]
\includegraphics{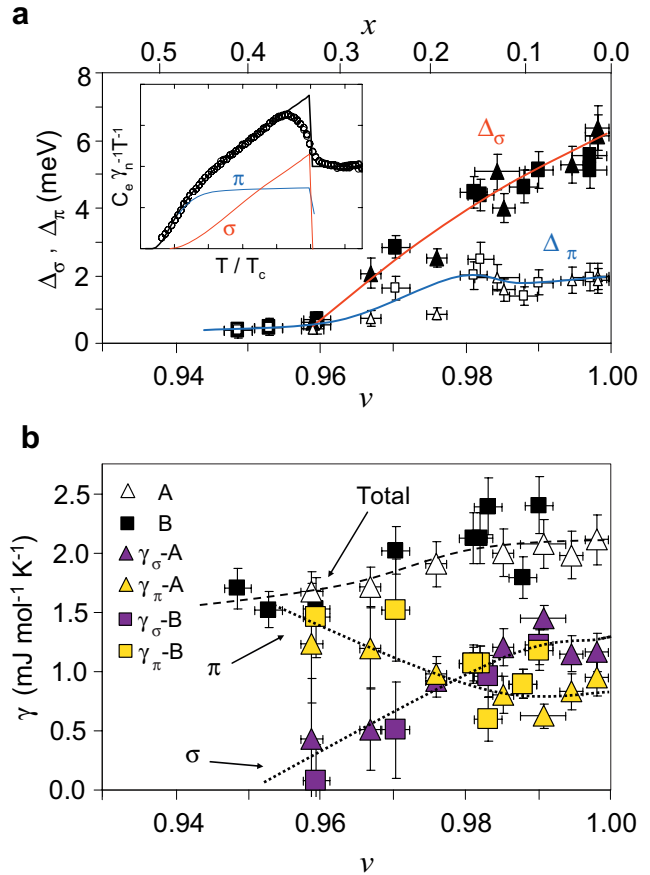}
\caption{The superconducting gaps for both sample sets extracted from heat capacity data are shown in plot (a).  The inset shows an example of the data obtained for a sample with $x \approx 0.2$.  Curves are guides to the eye.  The Sommerfeld coefficients extracted from the alpha model fitting are shown in plot (b).  Dotted lines are again guides to the eye.}
\label{f3}
\end{figure}

Heat capacity measurements were coordinated with the EELS analyses to monitor the superconducting gaps for the $\sigma$ and $\pi$ bands, $\Delta_\sigma$ and $\Delta_\pi$, and their DOS.  Heat capacity data were collected using a commercial system and analyzed with a 3-parameter fit (the so-called 'alpha' model \cite{alpha}) to extract the behavior of $\Delta_\sigma$ and $\Delta_\pi$ as a function of $v$.  Generally our data were very similar to those reported elsewhere \cite{Putti, Bouquet}.  

The results of the alpha-model fitting are shown in fig.~\ref{f3}a.  These data show a strong decrease of $\Delta_\sigma$ with decreasing $v$ (i.e. with higher Al content). The $\pi$ gap appears to hold constant for $v > 0.98$ and then decreases for smaller $v$.  However $\Delta_\pi$ remains well below $\Delta_\sigma$ until their values become equal at $v \approx 0.96$.  This merging point corresponds to the unit cell volume where the $T_c$ data changes curvature rather sharply.  As with the $T_c$ data in fig.~\ref{f1}, little difference is seen in the trends for samples made by either reaction A or reaction B, suggesting that these behavior are intrinsic to the average Al content in the grains.  The merging point of the gaps corresponds to $x = 0.33$ according to Vegard's law.    

For still smaller $v$, two gaps could not be distinguished, even though magnetization measurements demonstrate that bulk superconductivity survives, albeit only below 7 K.  These data extrapolate to zero gap for $v \approx 0.93$ ($x \approx 0.55$ from Vegard's law).  The alpha model also provides information about the Sommerfeld coefficients for the bands, which are proportional to the individual DOS.  Hence, the band responsible for this state can be inferred.  In fig.~\ref{f3}b, it can be seen that a progressive loss of the $\sigma$ band DOS occurs with decreasing $v$ (increasing $x$), extrapolating to zero for $v = 0.96$.  By contrast, the $\pi$ band DOS increases with $v$.  The $\sigma$ and $\pi$ DOS have approximately equal values at $v \approx 0.98$.  These trends suggest that it is the $\pi$ band that is the stronger band for $v < 0.96$, indicating a {\em reversal} of the 2-band hierarchy in that regime.  

We can summarize our results in terms of two important implications for understanding superconductivity in doped magnesium diborides.  First, our data are generally more consistent with the predictions of theory and less consistent with the existing experimental studies on Al-doped MgB$_2$.   In particular, a much stronger rate of decrease of $T_c$ and $\Delta_\sigma$ with composition (here derived from Vegard's law) has been found, indicating a stronger correlation between the loss of strong superconductivity on the $\sigma$ band and the topological transition of the Fermi surface \cite{Bianconi}.  $T_c$ falls from 39 to 7~K by the time $E_F$ covers the $\sigma$ band at the $\Gamma$ point in reciprocal space (at $x \approx 0.33$), confirming the loss of Fermi-surface area as the primary factor for $T_c$ in this region.  

Second, for $0.33 \alt x \alt 0.55$ the system is essentially a $\pi$ band superconductor with $T_c$ values below 7~K.  Superconductivity could be induced by the $\pi$ band onto the $\sigma$ band hole pockets, but their rapidly shrinking area makes their influence on the overall behavior small.  The complete loss of superconductivity at $x \approx 0.55$, when the $\sigma$ band is filled completely, correlates with the loss of the soft phonon modes in infrared absorption experiments \cite{Postarino,DiCastro}.  Thus, in this region the proximity of the $\sigma$ band to $E_F$ still provides interband electron-phonon coupling, even though the $\sigma$ band DOS does not strongly influence $T_c$.

Thus, the motion of the flat $\sigma$ band near the Fermi surface explains the key characteristics of superconductivity in Al doped MgB$_2$.  In this sense, superconductivity in MgB$_2$ can be thought of as being derived from stable AlB$_2$, where the $\pi$ bonding keeps the position of the $\pi$ band relatively fixed while the flat portion of the $\sigma$ band is moved up through $E_F$ by the addition of Mg.  An interesting question, therefore, is whether $T_c$ continues to increase as the $\sigma$ band moves upward away from $E_F$ for Na or Li doping (if it is possible).  Since the soft phonon modes should again stiffen as the $\sigma$ band moves higher, pure MgB$_2$ may lie close to the maximum of the $T_c$ vs.\ electron doping behavior.  Alternatively, it should be possible to take other diborides with stable $\pi$ bonding, such as ZrB$_2$, and alloy them with hole donors to drive them in the direction of a lattice instability to produce superconductivity, at least on the $\pi$ band.  

This work was supported by the U.S.\ Department of Energy, Office of Basic Energy Sciences under grant DE-AC02-98CH10886.  RFK acknowledges support as a Goldhaber Fellow at BNL.  LDC and AJZ acknowledge additional support from BNL-LDRD programs.  We would like to thank P. Canfield, A. Gurevich, J. Kortus, D. Larbalestier, W. Pickett, M. Suenaga, and D. Welch for stimulating discussions.


\begin{thebibliography}{99}
\bibitem{Suhl} H. Suhl, B. T. Matthias, and L. R. Walker, Phys. Rev. Lett. 3, 552–554 (1959).
\bibitem{Akimitsu} J. Nagamatsu, N. Nakagawa, T. Muranaka, Y. Zenitani, and J. Akimitsu, Nature 410, 63–64 (2001).
\bibitem{Mazin} I. I. Mazin and V. P. Antropov, Physica C 385, 49 (2003) and references therein.
\bibitem{Choi} H. J. Choi, D. Roundy, H. Sun, M. L. Cohen, and S. G. Loule, Nature (London) 418, 758 (2002).
\bibitem{Ahn}  J. An and W. E. Pickett, Phys. Rev. Lett. 86, 4366 (2001).
\bibitem{Yildirim} T. Yildirim et al., Phys. Rev. Lett. 87, 037001 (2001)
\bibitem{Slusky} J. S. Slusky et al., Nature vol.410, 343-5 (2001).
\bibitem{Postarino} P. Postorino et al., Phys. Rev. B 65, 020507 (2001).
\bibitem{Kazakov} S. M. Kazakov et al., Phys. Rev. B 71, 024533 (2005).
\bibitem{Ribeiro} R. A. Ribeiro, S. L. Bud'ko, C. Petrovic, and P. C. Canfield, Physica C 385, 16 (2003); W. Mickelson, J. Cumings, W. Q. Han, and A. Zettl, Phys. Rev. B 65, 052505 (2002).
\bibitem{Wilke} R. H. T. Wilke et al. Physical Review Letters vol.92, 217003(2004).
\bibitem{Luo} H. Luo, C. M. Li, H. M. Luo, and S. Y. Ding, J. Appl. Phys. 91, 7122 (2002).
\bibitem{Li-JQ} J. Q. Li et al., Phys. Rev. B 65, 132505 (2002).
\bibitem{Serventi} S. Serventi et al., Phys. Rev. B 67, 134518 (2003).
\bibitem{Papavassiliou} G. Papavissiliou et al., Phys. Rev. B 66, 140514 (2002).
\bibitem{Yang} H. D. Yang et al., Phys. Rev. B 68, 092505 (2003).
\bibitem{Putti} M. Putti, M. Affronte, P. Manfrinetti, and A. Palenzona, Phys. Rev. B 68, 094514 (2003).
\bibitem{delaPena} O. de la Pena, A. Aguayo, and R. de Coss, Phys. Rev. B 66, 012511 (2002).
\bibitem{Bussman-Holder} A. Bussmann-Holder and A. Bianconi, Phys. Rev. B. 67, 132509 (2003).
\bibitem{Ummarino} G. A. Ummarino, R. S. Gonnelli, S. Massidda, and A. Bianconi, Physica C 407, 121 (2004).
\bibitem{Badr} M. H. Badr and K.-W. Ng, Supercond. Sci. Technol. 16, 668 (2003).
\bibitem{Zandbergen} H. W. Zandbergen et al., Physica C 366, 221 (2002).
\bibitem{Egerton}R. F. Egerton, Electron Energy Loss Spectroscopy in the Electron Microscope,Plenum Press, New York, 1986.
\bibitem{Browning} N. D. Browning, M. F. Chisholm, and S. J. Pennycook, Nature 366, 143-146 (1993).
\bibitem{Klie} R. F. Klie et al., Phys. Rev. B 67, 144508 (2003).
\bibitem{Zhu} Y. Zhu et al., Phys. Rev. Lett. 88, 247002 (2002).
\bibitem{alpha} Y. Wang, T. Plackowski, and A. Junod, Physica C 355, 179 (2001).
\bibitem{Bouquet} F. Bouquet, R. A. Fisher, N. E. Phillips, D. G. Hinks, and J. D. Jorgensen, Phys. Rev. Lett. 87, 047001 (2002).
\bibitem{Bianconi} A. Bianconi et al., Phys. Rev. B 65, 174515 (2002).
\bibitem{DiCastro} D. Di Castro et al., Europhys. Lett. 58, 278 (2002).
\end{thebibliography}
\end{document}